\documentclass[%
 reprint,
 amsmath,amssymb,
 aip,
 cha,
floatfix,
]{revtex4-1}

\usepackage[T1]{fontenc}
\usepackage{graphicx}
\usepackage{dcolumn}
\usepackage{bm}
\usepackage[utf8]{inputenc}
\usepackage{mathptmx}
\usepackage{xcolor}

\newcommand{\im}{\text{i}}

\begin{document}


\title{Routes to spatiotemporal chaos in Kerr optical frequency combs}

\author{Aur\'elien Coillet}
\author{Yanne K. Chembo}%
 \email{yanne.chembo@femto-st.fr}
\affiliation{%
  FEMTO-ST Institute [CNRS UMR6174], Optics Department, \\
  16 Route de Gray, 25030 Besan\c con cedex, France.
}%


\date{\today}

\begin{abstract}
We investigate the various routes to spatiotemporal chaos Kerr optical frequency
combs obtained through pumping an ultra-high quality whispering-gallery mode resonator with a continuous-wave laser.
The Lugiato-Lefever model is used to build bifurcation diagrams with regards to the
parameters that are externally controllable, namely, the frequency and the power of the pumping laser.
We show that the spatiotemporal chaos emerging from Turing patterns and solitons display distinctive dynamical features.
Experimental spectra of chaotic Kerr combs are also presented for both cases, in excellent agreement with theoretical spectra.
\end{abstract}

\maketitle


\textbf{Optical Kerr frequency combs are sets of equidistant spectral lines generated
through pumping an ultra-high $Q$ whispering gallery mode resonator with a continuous wave laser~\cite{Kipp_2004,Maleki_PRL_LowThres,Kipp_Nature}.
The Kerr nonlinearity inherent to the bulk resonator induces a four-wave mixing (FWM) process, enhanced by 
the long lifetime of the intra-cavity photons which are trapped by total internal reflection in the ultra-low loss
medium. Four-wave mixing in this context allows for the creation and mixing of new frequencies as long as energy and momentum conservation laws are respected~\cite{YanneNanPRL,YanneNanPRA}. These Kerr combs are the spectral signatures of the dissipative spatiotemporal structures arising along the azimuthal direction of the disk-resonator. 
When the resonator is pumped in the anoumalous group-velocity dispersion regime, various spatiotemporal patterns can can build up, namely, Turing patterns, bright and dark solitons, breathers, or spatiotemporal 
chaos~\cite{IEEE_PJ,Arxiv_normal,Arxiv_anomalous}.
In this work, we investigate the evolution of the Kerr combs towards
spatiotemporal chaos when the frequency and the pump power of the laser are varied. 
We evidence the key bifurcations leading to these chaotic states and also discuss chaotic Kerr comb spectra obtained experimentally, which display excellent agreement with their numerical counterparts.} \\

\section{Introduction}

It was recently shown in refs.~\cite{Matsko_modelocked,PRA_Yanne-Curtis,Coen}
that Kerr comb generation can be efficiently modelled using the Lugiato-Lefever
equation~\cite{LL}, which is a nonlinear Schr\"odinger equation with damping,
detuning and driving. This partial differential equation describes the dynamics
of the complex envelope of the total field inside the cavity.

Spatiotemporal chaos is generally expected to arise in spatially extended
nonlinear systems when they are submitted to a strong excitation. In the context
of Kerr combs generation, such chaotic states have been evidenced theoretically
in ref.~\cite{YanneNanPRL} using a modal expansion method, which allowed to
demonstrate that the Lyapunov exponent is positive under certain circumstances.
The experimental evidence of these chaotic states was provided in the same
reference, and has been analyzed in other works as
well~\cite{Matsko_Chaos}. In the present article, we are focusing on
the routes leading to spatiotemporal chaos.  To the best of our knowledge, the
bifurcation scenario to chaos in this system is to a large extent unexplored,
despite the dynamical richness of this system.

\begin{figure}[b]
  \centerline{
    \includegraphics{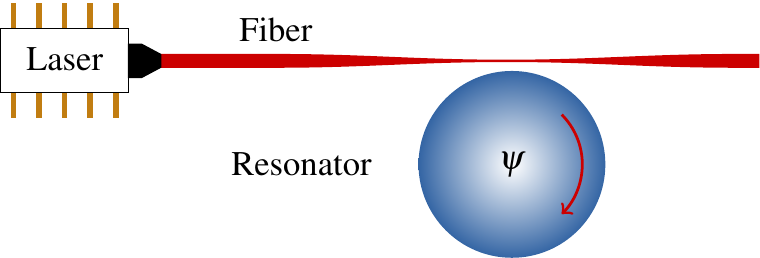}
  }
  \caption{Schematic representation of the Kerr comb generator. Continuous
    laser light is coupled into a ultra-high $Q$ whispering gallery-mode
    resonator using a tapered fiber. Inside the cavity, and above a certain threshold power, 
    the four-wave mixing induced by the Kerr nonlinearity can excite several eigenmodes, 
    thereby leading to a chaotic spatiotemporal distribution of the intra-cavity optical field.}
  \label{fig:Schema}
\end{figure}

The plan of the article is the following.
In the next section, we present the experimental system and the model used to
investigate its nonlinear dynamics.  A brief overview of the various dissipative
structures is presented in Sec.~\ref{dissipativestructures}.  The following
section is devoted to the route to chaos based on the destabilization of Turing
patterns, while Sec.~\ref{soliton} is focused on the second route which relies
on the destabilization of cavity solitons.  The experimental results are
presented in Sec.~\ref{experimental}, and the last section concludes the
article.

\begin{figure}[t]
  \centerline{
    \includegraphics{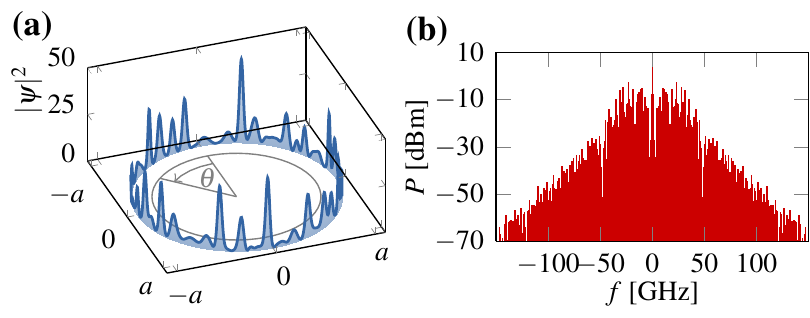}
  }
  \caption{(a) 3D representation of the simulated optical intensity inside the WGM
    resonator. In the chaotic regime, peaks of various amplitudes arise and disappear in the
    cavity. (b) Corresponding optical spectrum. This spectrum is made of discrete and quasi-equidistant 
    spectral lines, and for this reason, is generally referred to as  a Kerr optical frequency comb. } \label{fig:3DAndSpectrum}
\end{figure}

\section{The system}
\label{system}

The system under study consists of a sharply resonant whispering-gallery mode
cavity made of dielectric material, pumped with a continuous-wave laser (see
Fig. \ref{fig:Schema}).  The coupled resonator can be characterized by its
quality factor $Q=\omega_0/\Delta\omega$ where $\Delta\omega$ is the spectral
linewidth of the (loaded) resonance at the angular frequency $\omega_0$ of the
resonance.  Typical values for $Q$-factors leading to the generation of Kerr
combs are in the range of $10^9$.  In WGM resonators, the eigenmodes of the
fundamental family are quasi-equidistant, and these modes can be excited by the
pump excitation above a certain threshold. 

Kerr comb generation in high-$Q$ whispering-gallery mode resonators can be
described by two kinds of models. The first kind describes the evolution of each
resonant mode in the cavity, producing a system of $N$ ordinary differential
equations, with $N$ being the number of modes to be taken into
consideration~\cite{YanneNanPRL,YanneNanPRA}. While this modal model allows for
the easy determination of comb generation threshold, its computational cost
makes it inappropriate for studying the evolution towards chaos of the system.
Furthermore, only a finite, arbitrary number of modes can be simulated with this
model which reduces the accuracy of the numerical results, especially for
chaotic systems where a very large number of modes is excited.  While the modal
description consisted in large set of ordinary differential equations,
the spatiotemporal formalism provides an unique partial differential equation
ruling the dynamics of the full intracavity fieldrefs.~\cite{Matsko_modelocked,PRA_Yanne-Curtis,Coen}.
From the numerical point of view, this
latter description allows for easy and fast simulations using the split-step
Fourier algorithm, thus allowing the construction of bifurcation diagrams in
order to study the mechanisms that lead to chaos in a highly resonant and nonlinear optical
cavity.

In its normalized form, the LLE describing the spatiotemporal dynamics of the
intracavity field $\psi$ explicitly reads~\cite{PRA_Yanne-Curtis}  
\begin{equation}
  \frac{\partial\psi}{\partial \tau} = - (1 + i \alpha) \psi + i |\psi|^2 \psi -
  i \frac{\beta}{2}\frac{\partial^2\psi}{\partial \theta^2} + F 
  \label{eq:LLE}
\end{equation}
where $\tau = t/ 2\tau_{\rm ph}$ is a dimensionless time ($\tau_{\rm ph} = 1/\Delta\omega$ being the photon lifetime) and $\theta$ is the azimuthal angle along
the circumference of the disk-resonator.  The real-valued and dimensionless parameters of the
equation are the laser frequency detuning $\alpha$, the second-order dispersion
$\beta$, and the laser pump field $F$. The link between these dimensionless
parameters and the physical features of the real system are explicitly discussed
in refs.~\cite{PRA_Yanne-Curtis,IEEE_PJ}.

Chaos preferably arises in the system in the regime of anomalous dispersion, which
corresponds to $\beta<0$ (see refs.~\cite{YanneNanPRA,Arxiv_anomalous}).
It is also important to note that the two parameters
of the system that can be controlled experimentally are related to the pump
signal.  In particular, the laser pump power is proportional to $F^2$, and will
be the scanned parameter of our bifurcation diagrams. Intuitively, it can be understood that the more
energy is coupled inside the cavity, the more frequencies will be created
through the Kerr-induced four-wave mixing, ultimately leading to chaotic
behavior. The other parameter accessible to the experiments is the detuning
frequency $\alpha$ between the pump laser and the resonance. It is expressed in
terms of modal linewidths, in the sense that $\alpha = - 2(\Omega_0 - \omega_0)
/ \Delta \omega$ where $\Omega_0$ and $\omega_0$ are the laser and resonance
angular frequencies, respectively. 

At the experimental level, the most easily accessible characteristic of the
system is its optical spectrum. Effectively, while the LLE model rules the dynamics of the
spatiotemporal variable $\psi$, the spectrum can be obtained by performing the Fourier
transform of the optical field in the cavity. This duality is illustrated on
Fig.~\ref{fig:3DAndSpectrum}, and we will use it to compare our numerical
simulations to the experimental results.

\begin{figure}[tb]
\includegraphics{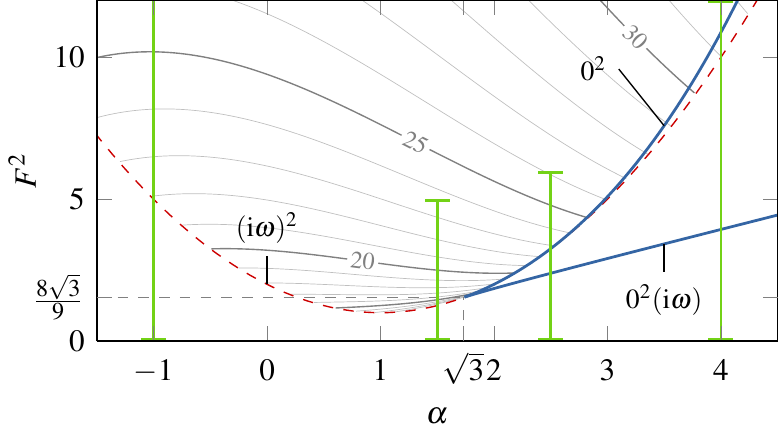}
\caption{\label{fig:StabMap} Stability diagram of the Lugiato-Lefever equation
for $\beta<0$ in the $\alpha$-$F^2$ plane~\cite{Arxiv_anomalous}. 
In between the two blue lines, three
flat solutions exist while only one is found outside this area. The red dashed
lines corresponds to the flat solution $|\psi|^2 =1$. 
Characteristic bifurcations can be associated to each of these curves. 
Note that the green lines vertical segments correspond to the bifurcation diagrams simulated in 
Figs.~\ref{fig:MI} and~\ref{fig:Soliton}. 
The thin gray curves indicate the number of rolls contained in the cavity when the system is in the Turing pattern
regime. }
\end{figure}

\section{Dissipative structures}
\label{dissipativestructures}

A stability analysis of the LLE in the $\alpha$-$F^2$ plane of parameters has
shown that various regimes of Kerr combs can be observed depending on the
detuning and pump power~\cite{Arxiv_anomalous}.
Figure \ref{fig:StabMap} represents this $\alpha$-$F^2$ plane where the spatial stability domain and
bifurcations lines of the stationary solutions have been drawn. This diagram
shows that the value $\alpha=\sqrt{3}$ is of particular interest, and we will
address the two cases $\alpha<\sqrt{3}$ and $\alpha>\sqrt{3}$ separately.

On the one hand, for detunings $\alpha$ below $\sqrt{3}$, the LLE only has one constant (``flat'')
solution for $\psi(\tau, \theta)$. It has previously  been shown in ref.~\cite{Arxiv_anomalous}
that this constant solution becomes (modulationally) unstable when the external pump $F^2$ reaches a
threshold value $F^2_{\rm th} = 1 + (\alpha -1)^2$, and Turing patterns (rolls) are formed.
Due to the periodicity of the variable $\theta$, only integer numbers of rolls can be found
in the cavity, and a close approximation of this number is given at threshold by
\begin{equation}
  l_{\rm th}=\sqrt{2(\alpha-2)/\beta}.
  \label{eq:NumberOfRolls}
\end{equation}
The iso-value lines for the rolls  in the $\alpha$-$F^2$
plane are displayed in the bifurcation diagram of Fig.~\ref{fig:StabMap}. 

On the other hand, when $\alpha$ is greater than $\sqrt{3}$, one to three constant solutions can be
found depending on the value of $F^2$. It is also in this area of the plane that
solitary waves can be found, such as cavity solitons or soliton breathers.
Unlike the previous case of Turing patterns, solitons always need an appropriate initial
condition to be excited.

In order to investigate the route towards chaos in this two cases, numerical
simulations were performed using the split-step Fourier method. The value for the loaded
quality factor was such that $Q = 10^9$, and a
slightly negative dispersion parameter was considered ($\beta=-0.0125$).
The evolution of the optical field in the cavity is simulated from the initial condition and for
a duration of $1$~ms, which is more than one hundred times longer than the photon lifetime $\tau_\text{ph} =
Q/\omega_0$. The extrema of the $100$ last intensity profiles are recorded and
plotted as a function of the excitation $F^2$, thereby yielding bifurcation diagrams. 
In the next sections, we present the results for the numerical simulations of the
LLE in these two different regimes.

\begin{figure}
\begin{center}
  \includegraphics{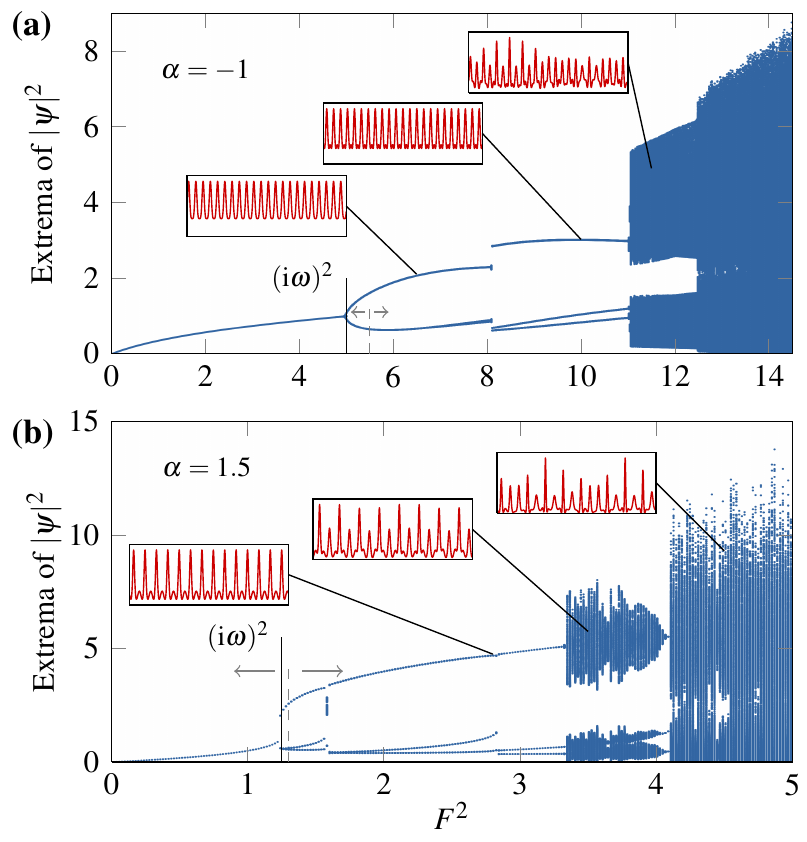}
\end{center}
\caption{\label{fig:MI} Bifurcation diagram in the case $\alpha=-1$ (a) and
$\alpha=1.5$ (b). In both cases, the initial condition is a small noisy background
just above the bifurcation $(\im\omega)^2$. This first bifurcation leads to the
formation of Turing patterns, number of rolls given by \ref{eq:NumberOfRolls}.
While the excitation is increased, these rolls become unstable and one or
several more rolls appear abruptly in the cavity. For higher gain, the
amplitudes of the peaks oscillate, and finally, a chaotic regime is reached.
Note that the inserts (in red) are snapshots displaying $|\psi|^2$ in a range $[-\pi, \pi]$.}
\end{figure}

\begin{figure}
\begin{center}
  \includegraphics{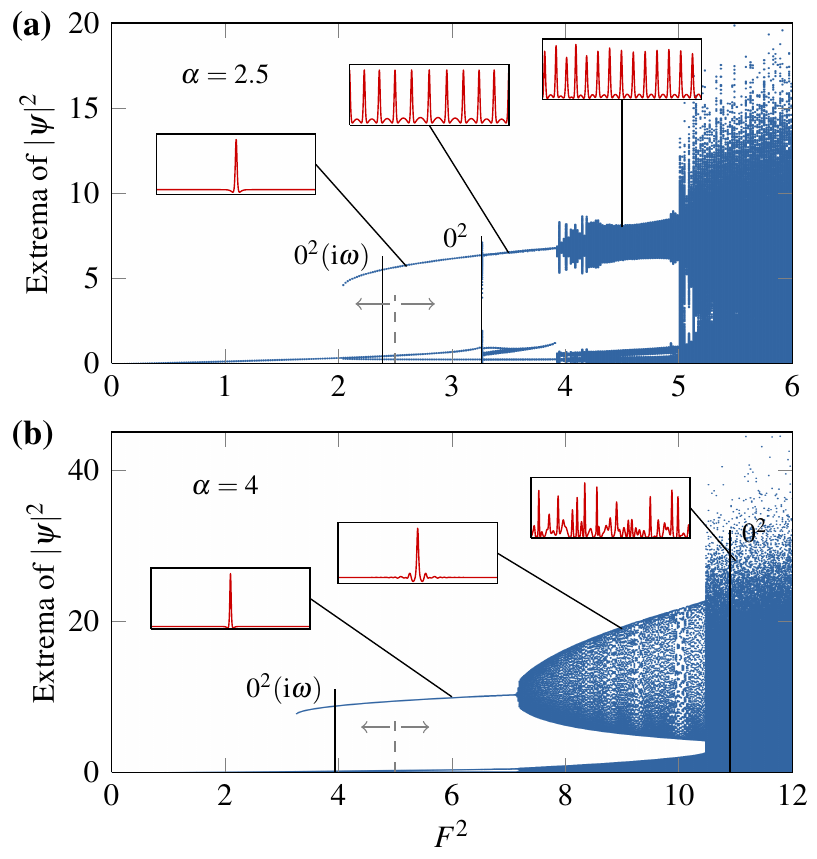}
\end{center}
\caption{\label{fig:Soliton} Bifurcation diagram in the cases $\alpha=2.5$ (a) and
$\alpha=4$ (b). In both cases the initial condition is chosen such that an unique
soliton is present in the cavity for excitations just above the $0^2(\im\omega)$
bifurcation. When the $0^2$ bifurcation is quickly reached (case $\alpha=2.5$),
the soliton remains stable until this bifurcation. At that point, other peaks
are created that fill entirely the cavity. For higher $F^2$, their amplitudes
vary in time, and finally, spatiotemporal chaos occurs. In the case $\alpha=4$,
the $0^2$ bifurcation happens at much higher excitations, and the soliton
becomes unstable and bifurcates to a breather characterized by a temporally
fluctuating amplitude. The system becomes chaotic in the vicinity of the $0^2$
bifurcation.
The inserts (in red) are snapshots displaying $|\psi|^2$ in a range $[-\pi, \pi]$.}
\end{figure}

\begin{figure}
\begin{center}
\includegraphics{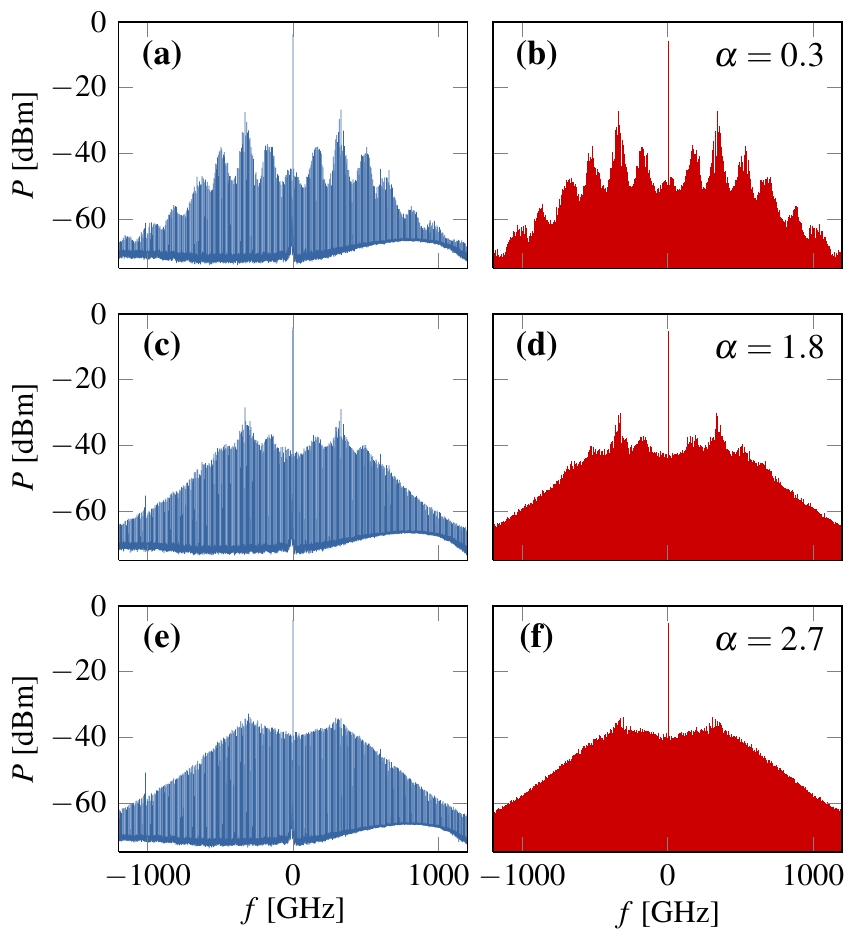} 
\end{center}
\caption{\label{fig:ExpVsNumSpectra} Experimental (left, a, c and e) and
simulated (right, b, d and f) Kerr combs obtained at same pump power and
different detunings. Excellent agreement is observed between experimental and theoretical results.
The frequency $f$ is relative to the laser frequency. This
frequency was decreased ($\alpha$ increased) between the spectra (a), (c) and
(e) while the pump power remained unchanged (around $300$~mW). The
numerical simulations were performed with $F^2=12$, $\beta=2.2\times10^{-3}$ and
$\alpha=0.3$ for the (b) case, $\alpha=1.8$ for (d) and $\alpha=2.7$ for (f).
While the first spectra (a) and (b) clearly arise from destabilized Turing patterns (note the spectral modulation),
the last sets spectra originate from the destabilization of cavity solitons.} 
\end{figure}

\section{Destabilization of Turing patterns}
\label{turing}

In the $\alpha<\sqrt{3}$ region of the bifurcation diagram, the starting point
of the bifurcation diagram simulations is a noisy background while the pump
power is above the threshold. In this condition, the flat solution is unstable
if the bifurcation diagram specifically starts at a value $\alpha < 41/30$,
and the noise allows the system to evolve toward stable rolls in the cavity.
After $1$~ms of simulation, the pump power is increased and the system is
simulated again, the previous final state being the initial condition. In order
to draw the lower part of the diagram, the same methodology is used, starting
from the same initial point. The resulting diagrams are presented in
Fig.~\ref{fig:MI}, for $\alpha=-1$ and $1.5$.

The $(\im \omega)^2$ bifurcation appears clearly in this diagrams, and
corresponds to the apparition of super-critical Turing patterns when $\alpha<41/30$
and sub-critical Turing patterns when $\alpha>41/30$ (see refs.~\cite{LL,IEEE_PJ,Arxiv_anomalous}).
At this point, the number of rolls in the cavity is given by Eq.~(\ref{eq:NumberOfRolls}) 
and Fig.~\ref{fig:StabMap} gives the values of $l$
above threshold. However, after further increase of the excitation, the number of rolls
in the cavity may not correspond to this ideal number, and the pattern changes
abruptly when the number of rolls has to shift.
This phenomenon is responsible for the discontinuities observed on the bifurcation diagrams. 

For very high excitation, the Turing patterns become unstable, starting with
oscillations in the amplitude of the rolls' peaks. These
fluctuations become stronger as $F^2$ is increased, finally leading to a
chaotic behavior with peaks of diverse amplitudes.

\section{Destabilization of solitons}
\label{soliton}

In the case $\alpha>\sqrt{3}$, the initial condition is chosen such that the
system converges to an unique soliton located at $\theta=0$. At this point, the
bifurcation diagram shows three extrema corresponding to the soliton maximum,
the continuous background and the dips on each side of the soliton (pedestals). 

Since the soliton is a sub-critical structure~\cite{PRA_Yanne-Curtis,LL,IEEE_PJ}
linked to a $0^2(\im \omega)$ bifurcation~\cite{Arxiv_anomalous}, it is still present for $F^2$ just
below the bifurcation value, as shown on Fig.~\ref{fig:Soliton}. When the
excitation is increased, the soliton maintains its qualitative shape, with the
continuous background being increased while the dips remains approximately at
the same level.

When the $0^2$ bifurcation is reached, the previously stable continuous
background becomes unstable, and the soliton can no longer be unique: a train of
identical pulses whose shapes are similar to the initial soliton appear in the
cavity. These pulses fill entirely the cavity, in a similar fashion to what was
observed in the $\alpha<\sqrt{3}$ case. 

For relatively small detunings, the subsequent evolution of the bifurcation
diagram is similar to the $\alpha<\sqrt{3}$ case, with fluctuations of the
pulses amplitudes ultimately leading to a chaotic repartition of the energy in
the cavity. For greater detunings ($\alpha=4$ in Fig~\ref{fig:Soliton}), the $0^2$
bifurcation happens at very high excitations, and the soliton remains alone in
the cavity for a large range of pump power $F^2$. This soliton becomes unstable in the
process, and transforms into a soliton breather whose peak amplitude
oscillates periodically with time. The amplitude of the oscillations increases with $F^2$
until the system explodes to chaos in the vicinity of the $0^2$ bifurcation.

\section{Experimental spectra}
\label{experimental}

In the experimental setup, a continuous-wave laser beam is amplified and coupled
to a WGM resonator using a bi-conical tapered fiber, as portrayed in Fig.~\ref{fig:Schema}.
The output signal is collected at the other end of the same
fiber and monitored either in the spectral domain with a high-resolution
spectrum analyzer, or in the temporal domain with a fast oscilloscope. It should
be noted that this output signal is the sum of both the continuous pump signal
and the optical field inside the cavity: the measured spectra will only differ
from the intra cavity spectra through an increased contribution from the pump laser.

The WGM resonator is made of magnesium fluoride (MgF$_2$) with refraction index $n_0 = 1.37$
at $1550$~nm. This dielectric bulk material has been chosen for 
two reasons. First, this crystal is characterized by very low
absorption losses, allowing for ultra-high $Q$-factors, equal to $2\times 10^9$ (intrinsic) in our case.
Secondly, the second-order dispersion of MgF$_2$ is anomalous at the
pumping telecom wavelength ($\lambda=1550$ nm), so that the waveguide dispersion is not
strong enough to change the regime dispersion; the overall dispersion generally remains anomalous
regardless of the coupling.
The diameter of the WGM resonator is $d \sim 11.3$~mm and its
free-spectral range (FSR -- also referred to as \textit{intermodal frequency})
is $\Delta \omega_{\rm FSR}/2 \pi = c/  \pi n_0 d =5.8$~GHz,
where $c$ is the velocity of light in vacuum.

Once an efficient coupling is achieved, the pump power is increased to few hundreds
of milliwatts, and the detuning is adjusted so that Kerr frequency combs
are generated. Further adjustments of these two parameters allow us to reach
chaotic regimes where every mode of the fundamental family in the WGM resonator is populated.
In the example of Figs.~\ref{fig:ExpVsNumSpectra}(a), (c) and (e), the pump power was
fixed and the detuning was changed from a negative value to a positive value by
increasing the wavelength. The bump in the background noise level is due to the unfiltered
amplified spontaneous emission (ASE) originating from the optical amplifier used in the
experiment. The comb (a) corresponds to a destabilized Turing pattern where the
so-called primary and secondary combs~\cite{YanneNanPRA} have evolved by
populating nearby modes until every mode is filled. On the contrary, the
structure of spectrum (e) is smooth, and originates from the destabilization of cavity
solitons. The comb presented in (c) stands in the transition
between the two different regimes. In each case however, important fluctuations
of the amplitude of each mode are observed, confirming the chaotic nature of
these experimental Kerr combs. To obtain representative spectra, the (a), (c) and (e)
spectra were averaged over 30 consecutive measurements.

In Fig.~\ref{fig:ExpVsNumSpectra}, we also provide numerical simulations of the
optical spectra that qualitatively correspond to their experimental
counterparts. The dispersion parameter was determined using
Eq.~(\ref{eq:NumberOfRolls}) and an experimental spectra with Turing patterns near
threshold, yielding $\beta=2.2\times10^{-3}$.  

Similarly to the experiments, the simulated combs were obtained for a fixed pump
power $F^2=12$ and three different detunings. The first one, $\alpha=0$
correspond to the Turing pattern case while the last one $\alpha=2.7$ is in the
soliton regime. The transition case was taken for $\alpha=1.8$, close to the
$\alpha=\sqrt{3}$ limit. As stated previously, the power of the central mode has
been increased with $11.5$~dBm to account for the differences between the
intra-cavity and output spectra. To simulate the
integration of the spectrum analyzer, the spectra are averaged over the last
$1000$~iterations. The agreement between numerical simulations and experiments is excellent: it covers an impressive
dynamical range of $80$~dB, and a spectral range of $2$~THz which corresponds to more than $300$ modes. 
This agreement also validates the interpretation of these combs as the result of two different
routes to spatiotemporal chaos.

\section{Conclusion}

In this work, we have investigated the generation of optical Kerr combs in WGM
resonators using the Lugiato-Lefever equation to describe our experimental
system. Numerical simulations allowed us to draw bifurcation diagrams of the
system for various frequency detuning values. The natural scanning parameter in this case is
the pump power injected in the resonator. Depending on the value of the detuning
parameter, various routes to chaos have been evidenced. In particular, for
detunings $\alpha$ below $\sqrt{3}$, the chaotic Kerr combs originate from
destabilized Turing patterns. In this case, the envelope of the corresponding
Kerr comb displays several maxima which are reminiscent of the primary and
secondary combs encountered at lower excitation. For $\alpha$ higher than
$\sqrt{3}$, the evolution of a soliton can either lead to a set of unstable
pulses filling the cavity or optical breathers before reaching chaos. The
optical spectrum in this case is smooth, with a shape similar to the soliton's
spectrum. In both cases, the numerical spectra are in excellent agreement with
the experimental results obtained using a ultra-high $Q$ magnesium fluoride resonator, thereby
providing a strong validation of the model and simulations. \\

\begin{acknowledgments}
The authors acknowledge financial support from the European Research Council
through the project NextPhase (ERC StG 278616).
\end{acknowledgments}


\end{document}